\documentclass[article,aps,twocolumn]{revtex4-1}
\usepackage[latin1]{inputenc}
\usepackage[english]{babel}
\usepackage{color}
\usepackage{graphicx}
\usepackage{ulem}

\begin{document}
\title{Subcritical temperature in Bose\,-\,Einstein condensates of chiral molecules}
\author{Pedro Bargue\~no $^{1,2}$\footnote[1]{pbdr@imaff.cfmac.csic.es},
 Ricardo P\'erez de Tudela $^1$ \footnote[3]{rpt@imaff.cfmac.csic.es},
Salvador Miret-Art\'es$^1$\footnote[4]{s.miret@imaff.cfmac.csic.es}
and Isabel Gonzalo $^3$\footnote[2]{igonzalo@fis.ucm.es}
} \affiliation{ $^1$ Instituto de F\'{\i}sica Fundamental (CSIC),
Serrano 123, 28006 Madrid, Spain
\\
$^2$ Departamento de Qu\'{\i}mica F\'{\i}sica, Universidad de
Salamanca, 37008 Salamanca, Spain
\\
$^3$ Departamento de \'Optica, Universidad Complutense de Madrid,
28040 Madrid, Spain }

\begin{abstract}
Bose-Einstein condensation of a sample of non-interacting chiral
molecules leads to a non zero optical activity of the 
condensate and also to a subcritical temperature in the heat 
capacity. This is due to the internal structure of the molecule 
which, in our model, is considered as a simple two-state system,
characterized by tunneling and parity violation.
The predicted singular behavior found for the specific heat, below 
the condensation temperature, sheds some light on the existence of the
so far elusive parity violation energy difference between enantiomers.
\end{abstract}

\maketitle


Since the prediction and subsequent discovery of parity 
violation \cite{Lee1956,Wu1957} in weak interactions, the role of
discrete symmetry breaking in fundamental physics is an intriguing
field of research. Although the weak interaction (between electrons
and nuclei) mediated by the gauge $Z^0$ boson has been extensively
studied and observed in atoms \cite{Bouchiat1997}, it has only been
predicted in molecules. 
The importance of this 
symmetry-breaking in molecules is twofold: (i) it 
could be intimately related to the origin of
homochirality, that is, the almost exclusive one-handedness of the
chiral molecules found in living systems, 
this being one of the most fascinating open problems which
links  fundamental physics  with the biochemistry of life
\cite{Guijarro2009} and (ii) at fundamental level, the intrinsic chiral nature which is present 
in some molecules should reflect the underlying interaction containing pseudoscalar magnitudes as those 
appearing in the weak one. In molecular systems, the theory of
electroweak interactions predicts a parity violating  energy
difference (PVED) between the two enantiomers of chiral molecules
to be between $10^{-13}$ and $10^{-21}$ eV
\cite{Zanasi99,Laerdahl00,Thyssen2000,Lahamer2000,Soulard06,Figgen2008,
Quack2008}.
However, no conclusive energy difference has been reported, for
example, in experimental spectroscopic studies of the CHBrClF
molecule reaching an energy resolution of about $10^{-15}$ eV
\cite{Daussy99,Crass05}. 

Due to the fact that the parity violating signals are easily 
masked by thermal effects, it is highly desirable to reach cold 
or ultracold regimes in the laboratory \cite{Flambaum06,Blythe2005,Hudson06,Hudson07}.
The measurement of fundamental physical properties such as the
PVED between enantiomers could be achieved by trapping
molecules at low temperatures (in the milikelvin range or below) and,
subsequently, performing ultrahigh-resolution spectroscopic
measurements of vibrational or electronic transitions. 
We would like to point out that these temperatures, and even colder,
have been reached
within the actual experimental capabilities only for atoms and diatomic molecules. 
The rapid expansion of the research field of ultracold chemistry
is already opening new and exciting possibilities concerning more complex systems
\cite{Carr2009}. Thus, the study of quantum thermal
effects in chiral molecules through pseudoscalar 
operators as the basic
object describing their thermodynamics is of fundamental 
importance. Very recently, we have studied the corresponding
classical thermodynamics of non-interacting chiral molecules 
\cite{BarguenoCPL2009,GonzaloCPL2010}.
In this sense, as pointed out by Flambaum and coworkers concerning some capabilities of
Bose-Einstein condensates to amplify weak interactions in chiral molecules \cite{Flambaum06},
the study of Bose-Einstein condensation (BEC) for a gas of chiral molecules could shed some light
on some properties relevant to determine the so far elusive PVED. 

On the other hand, the study of cold atoms
with intrinsic spin structure is being a very hot topic, including
two-component condensates, spinor condensates, etc \cite{Pethickbook}. 
In this sense, for a Bose gas of chiral molecules, 
the inclusion of extra information about the internal structure 
of such molecules could be determinant to obtain some
macroscopic properties relevant to determine the PVED such as,
for example, by measuring the optical activity and heat capacity. 
For this end, we will employ a two-state model which
includes tunneling and parity violating effects in chiral molecules (this approach was first considered
in \cite{Harris78} and is widely used to describe, for instance, the internal dynamics of chiral molecules).
Although the two-state (or spin 1/2) model is well
known in many fields, including spin tunneling, magnetization,
quantum dissipative systems, to name only a few, this is the first time, to the best of our
knowledge, that this model is employed to study BEC of chiral molecules.
Following the seminal works on BEC, predicted for a non-interacting gas 
and carried out in numerous dilute atomic gases \cite{Anderson1995} and in some diatomic
molecules as in $^{40}\mathrm{K}_{2}$ \cite{Regal2003}, we will assume that
the interaction among molecules is negligible at first order and, 
therefore,  a non-interacting quantum gas of chiral molecules will be
considered.

In this Letter we 
show that a condensed gas of non-interacting chiral molecules
displays optical activity. In addition, we show 
a dramatic change of the heat capacity with temperature, leading 
to the appearance of a secondary maximum (developing
a shoulder-type-structure) associated with the Schottky anomaly
below the condensation temperature. This {\it subcritical} temperature is expected
to lie in the cold or ultracold regime.

In an ideal Bose-Einstein gas, only the translational part of the
molecular motion is involved and the critical temperature
for BEC is reached when the fugacity of the gas $z = \exp({\beta \mu})$,
with $\mu$ being the chemical potential and $\beta = (k_{B}T)^{-1}$, 
is approaching unity (assuming that the 
energy of the ground state is zero) \cite{Feynman1998}.
This $T_c$ depends on the molecule mass and the
density of the system. At $T < T_c$, there must be a non-negligible
fraction of molecules in the ground state, the internal energy goes
with $T^{5/2}$ and the heat capacity  with $T^{3/2}$. If each chiral
molecule in the condensed phase has its own internal structure, we
could ask ourselves if the corresponding quantum statistics describing
the thermodynamical behavior, intimately related to the PVED,
can be manifested by a macroscopic effect. For this goal, a two-state
model is assumed for each chiral molecule.

In the absence of parity violation, the dynamics of a chiral
molecule is well described by a symmetric double well potential.
In this case, the true stationary states of a chiral molecule
are the (achiral) eigenstates of parity. The effect of including
in the Hamiltonian the internal P-odd term given by the parity
violating electron-nucleon electroweak interaction, $H^{PV}$,
introduces an energy difference between the two minima of
the double well potential, leading to a new set of energy
eigenstates.
Let us consider the total Hamiltonian of the system
$H=H^{0}+H^{PV}$, with $H^{0}$ including only parity conserving
terms. Using a two-state model, we choose the left and right
chiral states basis, $|L\rangle$ and $|R \rangle$ (localized
respectively in the left and right minimum of the double well
potential), in order to show clearly the parity properties of the
Hamiltonian \cite{Harris78}:
\begin{equation}
\label{eq1}
H = H^0 + H^{PV} = \delta \sigma_{x} + \epsilon_{PV}\sigma_{z}
\end{equation}
where $\sigma_{x,z}$ are the Pauli matrices.
The eigenstates of $H$, $|1,2\rangle$, can be
expressed as linear combinations of the chiral states by means of a rotation 
where the corresponding mixing angle, $\theta$, obtained from the knowledge of
the eigenvectors of $H$, is given by 
$\tan 2\theta=(2\, H_{LR} / H_{RR}-H_{LL})=
( \delta / \epsilon_{PV})$.
The energy splitting between the two eigenstates of
$H^{0}$ is $2 \delta>0$, where $\delta= \langle L|H^{0}|R\rangle$.
This magnitude is related to the height of the barrier of the
double well potential 
and is inversely proportional
to the tunneling time. The PVED is then given by
$|H_{RR}-H_{LL}|= |2\epsilon_{PV}|$ with $\epsilon_{PV}=\langle
R|H^{PV}|R\rangle= -\langle L|H^{PV}|L\rangle $ (we remark that $\epsilon_{PV}$ is the eigenvalue of a 
pseudoscalar operator in the chiral basis). 
The eigenvalues
of the system are given by $E_{1,2}=E_{0}\mp \Delta$, with
$E_{0}=(H_{RR}+H_{LL})/2$ and $\Delta \equiv \sqrt {\epsilon^2_{PV}+\delta^{2}}$
(hereafter we will take $E_{0}=0$ for the sake of simplicity).

{\it Bose-Einstein condensation of a non-kinetic-two-state gas}.
In a gas of non interacting chiral molecules at
temperature T, the inclusion of P-odd effects leads to deal
with a pseudoscalar operator for obtaining any thermodynamic variable \cite{BarguenoCPL2009,GonzaloCPL2010}.
Thus, for a system described by a biased double well potential, 
the only magnitude
which, roughly speaking, distinguishes between left and right
conformations is the population difference between both wells,
$N_{L}-N_{R}$, which is directly related to the optical activity
of the system. If $X$ is a pseudoscalar operator, it can be
shown that
\begin{equation}
\label{eqfin} |\langle X \rangle_{\mathrm{stat}}| = 
|x| \, (N_{L}-N_{R})_{\mathrm{stat}} ,
\end{equation}
where $\pm x$ are the eigenvalues of $X$ and the subscript
{\it stat} stands for the different statistics considered.
Eq. (\ref{eqfin}) is a generalization to any statistics
of the Maxwell-Bolztmann (MB) thermal average \cite{BarguenoCPL2009}.
A straightforward calculation gives the expression for
$N_{2}-N_{1}$ to be
\begin{equation}
\label{eqn}
(N_{2}-N_{1})_{\mathrm{stat}}=\tanh \beta \Delta  \,(1+p \, \frac{z+z^{-1}}{2}\, 
\mathrm{sech}\beta \Delta)^{-1},
\end{equation}
where
$p=0$ applies for MB, $p=-1$ for Bose-Einstein (BE) and
$p=+1$ for Fermi-Dirac statistics. We remark that, when $\epsilon_{PV}$ dominates over $\delta$, the
true eigenstates of the system are the chiral states, $|L\rangle$ and $|R\rangle$, and changing
the basis from ${|1\rangle ,|2\rangle}$ to ${|L\rangle ,|R\rangle}$ introduces the factor
$\cos 2 \theta = \epsilon_{PV} / \Delta$ leading to 
$N_{L}-N_{R}= \epsilon_{PV}(N_{2}-N_{1}) / \Delta$.
Moreover, $N_{L}-N_{R}$ changes its sign when considering a
parity-transformed double
well since $\epsilon_{PV}$ changes to $-\epsilon_{PV}$.
From the knowledge of $N_{2}-N_{1}$, the internal energy
of the gas is 
\begin{equation}
\label{eqU}
U = -\Delta (N_{1}-N_{2})_{\mathrm{stat}}.
\end{equation}
From Eq. (\ref{eqU}) one can inmediately obtain the heat capacity 
at constant volume. This $C_v$ can be interpreted as
a measure of the fluctuations of the optical activity of the system,
that is, a measure of the fluctuations of the pseudoscalar character 
of the system.

Let us now consider BE statistics ($p=-1$) for a non-kinetic-two-state
gas, that is, the Hamiltonian $H$ includes tunneling
and parity violation but not kinetic terms. The total number of
particles is $N=N_{1}+N_{2}$, where $N_{1,2}$ is the number of
particles in the ground, $|1\rangle$,
and the excited state, $|2\rangle$, respectively. Using the standard
expression for the occupation numbers for  Bose statistics, we get 
$N=N_{1}+ (z^{-1}\exp (\beta \Delta)-1)^{-1}$. At the condensation 
temperature, $T^*$, all the particles are in the excited state, $N_{2}$.
(this definition of the condensation temperature was also pointed out  
in \cite{Kett1996}). This leads to the condensation temperature to
be determined from
\begin{equation}
T^*=\frac{2\Delta}{k_{B}\ln (1+\frac{1}{N})}.
\end{equation}

The condensate displays a non-zero optical activity which is proportional to
\begin{equation}
N_{2}-N_{1}=1-2 \frac{\exp (2 \beta^* \Delta)-1}{\exp (2 \beta \Delta)-1},
\end{equation}
and this enables us to write, for $T<T^*$:
\begin{equation}
\frac{N_{0}}{N}=1-\frac{\exp (2 \beta^* \Delta)-1}{\exp (2 \beta\Delta)-1},
\end{equation}
the average internal energy as
\begin{equation}
u= \frac{U}{N}= \Delta(2\frac{\exp (2 \beta^* \Delta)-1}{\exp (2 \beta \Delta)-1}-1),
\end{equation}
and the heat capacity as
\begin{equation}
\label{cvbec}
C_{v}=\frac{\partial u}{\partial T}=4 k_{B} (\beta \Delta)^{2} \frac{\exp (2 \beta^* \Delta)}
{\exp (2 \beta \Delta)-1}.
\end{equation}
It is worth noting that this heat capacity reaches its maximum value at
the {\it subcritical} temperature, $T_{sc}$, 
\begin{equation}
\label{tsc}
\beta_{sc}\Delta \simeq 0.797
\end{equation}
which reflects nothing but the Schottky anomaly due to the saturation
of the energy levels of the system.

\begin{figure}[h*]
\begin{center}
\includegraphics[angle=-90,width=0.43\textwidth]{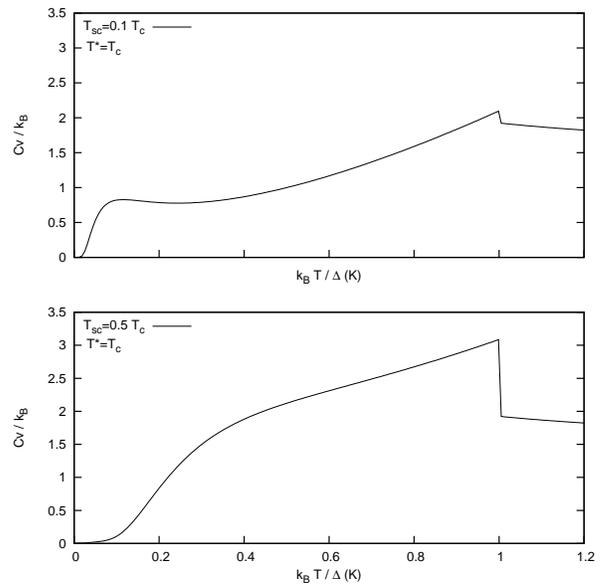}
\caption{Total heat \label{fig1} capacities for a gas of
chiral molecules as a function of the reduced temperature
for two cases: $T_{sc} = 0.1 T_c$ and $T_{sc} = 0.5 T_c$. 
The gas undergoes BEC at $k_{B} T/\Delta$ = 1. We have taken $T^*=T_{c}$. See text for symbols.}
\end{center}
\end{figure}

{\it Bose-Einstein condensation of a kinetic-two-state gas}. 
If we include a kinetic
term in the Hamiltonian, the total partition function for an ideal Bose 
gas of chiral molecules in the two-state model can be written as 
\begin{equation}
Z_{tot}=Z_{kin}\cdot Z_{int}
\end{equation}
where the subscripts denote the kinetic and internal contributions,
respectively. The first factor can be found in any standard textbook
of Statistical Mechanics \cite{Feynman1998} and the second factor
is given, for our case, by
\begin{equation}
\label{zint}
Z_{int} = \prod_{i=1,2}(1-z \, e^{-\beta E_{i}})^{-1}.
\end{equation}
The factorization of the partition function leads to a sum of the 
corresponding heat capacities, $C_{v}^{tot} = C_v^{kin} + C_v^{int}$. 
We note that there are now two temperature regimes involved in the
chiral system: (i) the temperature at which the gas undergoes
BEC ($T_c$) and (ii) the {\it subcritical} temperature ($T_{sc}$, 
below $T_c$) where the heat capacity displays a maximum. 
In the case of a free gas, $T_c$ is related to the mass and number
density of the bosons but $T_{sc}$ depends only (under the two-state
model) on the energy splitting between $|1 \rangle$ and $|2 \rangle$ states. Thus, for $T<T_{c}$, 
$C_v^{int}$ is given by Eq. (\ref{cvbec}) and, for $T>T_{c}$, 
the corresponding expression can be obtained from Eq. (\ref{eqU}). $C_v^{kin}$ is the usual heat 
capacity for an ideal Bose gas. It is worth pointing out
that the very interesting case where $T_c$ and $T_{sc}$ are 
of the same order gives
place to an appreciable change in the heat capacity and 
to a shoulder-type structure
of it below $T_c$. This behavior is displayed in Fig. \ref{fig1}, 
where the total heat capacity is plotted in terms of the reduced temperature $k_{B} T / \Delta$ for 
the cases $T_{sc} = 0.1 T_c$ (top panel) and $T_{sc} = 0.5 T_c$ (bottom panel) (we have assumed $T^*=T_{c}$). 

Strictly speaking, the effect of $\Delta$ on the condensation
temperature, $T_c$, results into a very small shift. Thus 
for a kinetic-two-state-gas we get that the corrected or shifted
condensation temperature is given by
\begin{equation}
\Upsilon = T_c \left ( 1- \frac{\mathrm{cosech} \beta^* \Delta}
{\mathrm{sech} \beta^* \Delta + e^{-3 \beta^* \Delta} -1}  \right ) .
\end{equation}
This difference is so tiny that for most practical purposes, we
keep $T_c$ instead of $\Upsilon$.

When $T_c=T_{sc}$,  the subcritical temperature corresponds to a very
precise density of bosons of mass $m$, $n_{sc}$, 
given by
\begin{equation}
\label{critdens}
n_{sc} \simeq 0.178 \left( \frac{m \Delta}{\hbar^2} \right)^{3/2} .
\end{equation}
The bottom panel of Fig. (\ref{fig1}) should be observed, for example,
for the T$_2$Se$_2$ system. This system has $\epsilon_{PV}\approx$ 2.5$\cdot10^{-14}$ eV, 
$\delta\approx$ 5$\cdot10^{-17}$ eV and the
tunneling time is $\sim$ 40 s (see Table 2 of \cite{Quack2008}). 
Its subcritical density, $n_{sc}$, is about 
$10^{10}$ cm$^{-3}$, which is of the order of the critical densities. 
Thus, in this case, the corresponding discontinuity in the heat
capacity could provide us a direct and clear signal of molecular parity
violation.  

\begin{figure}[h*]
\begin{center}
\includegraphics[angle=0,width=0.5\textwidth]{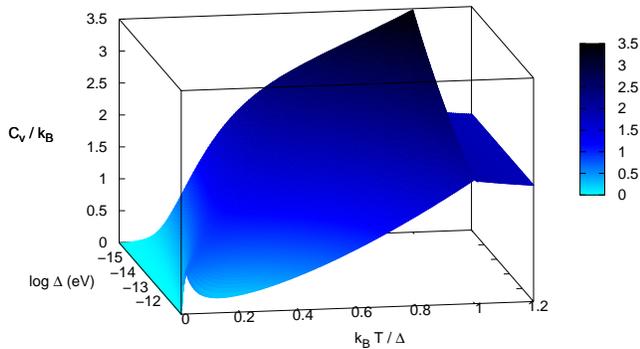}
\caption{\label{fig2} Three dimensional plot of the heat capacity
versus $\Delta$ and the reduced temperature. The gas undergoes BEC at $k_{B}T/\Delta =1$.
We have taken $T^*=T_{c}$. See text for symbols.}
\end{center}
\end{figure}

In Fig. (\ref{fig2}), a three dimensional plot of the heat capacity
in the range of variation $[0,1.2]$ of the reduced temperature and
$[10^{-16},10^{-11}]$ of the splitting $\Delta$ (eV) is showed. In this plot it is clearly seen the overall 
shoulder-type structure as well as the subcritical maximum for different chiral molecules, characterized
by the splitting $\Delta$. This splitting is representative of molecules such as, for example, H$_2$S$_2$, 
CHBrClF and H$_2$Se$_2$ \cite{Quack2008}.

The subcritical maximum (or anomalous shoulder-type structure)
found here could be considered as a signal of the Schottky anomaly
which comes from lifting the degeneracy of the internal degrees of freedom. Although this was previously 
considered in the framework of a path integral approach to 
study BEC in spinor gases \cite{spinorPRE}, our simple model allow us to identify more clearly
the anomaly.
We would like to stress that the {\it subcritical}
temperature is determined in essence by $\delta$ and
$\epsilon_{PV}$, these temperatures lying in the cold or ultracold
regime. The existence of that temperature could be possible for any
system
whose constituents have internal ground states slightly splitted,
such as for example: the inversion doubling of non-rigid pyramidal
molecules, the splitting due to torsional or internal rotations
through potential barriers in non-rigid molecules, or eventually
the hyperfine structure of the ground state of atoms and molecules 
(in this last case, it could be more easy to verify the type of
anomaly here studied). However, chiral molecules in this two-state 
model play a special role in BEC when the splitting is mainly due to parity violation
(as in T$_2$Se$_2$). In this case, an observation of the anomaly predicted in the heat
capacity provides a direct information about the PVED.
In addition, when $\epsilon_{PV}$ determines the dynamics, then
the eigenstates of the system tend to be the chiral states $|L,R\rangle$. 
We also point out that 
a measurement of the optical activity of the condensate would be a confirmation of
the existence of the PVED, noting that the factor $\epsilon_{PV}/\Delta$, which is the
optical activity for zero temperature, should be 
measurable by state-of-the-art of actual polarimeters, as we concluded in \cite{GonzaloCPL2010}. 

Finally, we would like to point out that
the energy scale of molecular parity violation is associated with 
the natural scale of densities  
which corresponds to those achieved in Bose-Einstein condensates. 
Thus, BEC seems to be essential to detect the PVED.
Furthermore, as is it well known, interactions could modify the physics of the problem
so it is fundamental to ask ourselves to what extent the anomaly in the heat capacity persists when we
consider an interacting system of chiral molecules. Although one could introduce the interaction between molecules 
under the Gross-Pitaevskii or any other more sophisticated treatment, an alternative way is to study the
molecular sample as an open quantum system under the influence of dissipation. 
In this sense, we note that the heat capacity 
anomalies of open quantum systems have been recently studied when coupled to a thermal bath \cite{Hanggi2009},
showing the robustness of these anomalies when the interaction is taken into account. Thus, we expect that
a similar stability for the anomaly here considered will persist even in presence of an environment. 
The study of heat capacity anomalies in BEC chiral gases embedded in appropiately chosen environments is 
currently in progress.

\vspace{0.5cm}

This work has been funded by the MEC (Spain) under projects
CTQ2008-02578/BQU, FIS2007-62006 and FIS2007-65382, 
supported by grants BES-2006-11976 (P. B.) and BES-2006-7454 (R. P. de T.).
P. B. dedicates this work to Ana\'{\i}s Dorta-Urra for her help and encouragement during
last months.

\end{document}